\documentclass[aps,aps, llncs, preprint]{revtex4}
\usepackage{amsmath}
\usepackage{amsfonts}
\usepackage{amssymb}
\usepackage{graphicx}
\usepackage{amsmath}
\usepackage{float}
\usepackage{color}
\usepackage{soul}
\usepackage[utf8]{inputenc}
\usepackage[T1]{fontenc}

\DeclareUnicodeCharacter{0096}{_}

\begin{document}
\title{Superconductivity coexisting with ferromagnetism in quasi-one dimensional non-centrosymmetric (TaSe$_4$)$_3$I} 

\author{Arnab Bera$^1$, Sirshendu Gayen$^{1,2}$, Suchanda Mondal$^3$, Riju Pal$^4$, Buddhadeb Pal$^4$, Aastha Vasdev$^5$, Sandeep Howlader$^5$, Manish Jana$^1$, Tanmay Maiti$^3$, Rafikul Ali Saha$^1$, Biswajit Das$^1$, Biswarup Satpati$^3$, Atindra Nath Pal$^4$} 

\email{atin@bose.res.in}

\author{Prabhat Mandal$^{3,4}$}
\email{prabhat.mandal@bose.res.in}

\author{Goutam Sheet$^5$}
\email{goutam@iisermohali.ac.in}

\author{Mintu Mondal$^1$}
\affiliation{$^1$Indian Association for the Cultivation of Science, Jadavpur, Kolkata 700032, India.}
\affiliation{$^2$Department of Physics, Chandigarh University, Gharuan, Mohali, Punjab 140413, India.}
\affiliation{$^3$Saha Institute of Nuclear Physics, HBNI, 1/AF Bidhannagar, Calcutta 700 064, India.}
\affiliation{$^4$S. N. Bose National Centre for Basic Sciences, Sector III, Block JD, Salt Lake, Kolkata 700106, India.}
\affiliation{$^5$Indian Institute of Science Education and Research (IISER) Mohali, Sector 81, S. A. S. Nagar, Manauli, PO 140306, India.}

\date{\today}

\begin{abstract}
\textbf{Low-dimensional materials with broken inversion symmetry and strong spin-orbit coupling can give rise to fascinating quantum phases and phase transitions. Here we report coexistence of superconductivity and ferromagnetism below 2.5\,K in the quasi-one dimensional crystals of non-centrosymmetric (TaSe$_4$)$_3$I (space group: $P\bar{4}2_1c$). The unique phase is a direct consequence of inversion symmetry breaking as the same material also stabilizes in a centro-symmetric structure (space group: $P4/mnc$) where it behaves like a non-magnetic insulator\cite{cTSI1, cTSI2, cTSI3, NbSe4_2I_prb88}. The coexistence here upfront contradicts the popular belief that superconductivity and ferromagnetism are two apparently antagonistic phenomena. Notably, here, for the first time, we have clearly detected Meissner effect in the superconducting state despite the coexisting ferromagnetic order. The coexistence of superconductivity and ferromagnetism projects non-centrosymmetric (TaSe$_4$)$_3$I as a host for complex ground states of quantum matter including possible unconventional superconductivity with elusive spin-triplet pairing\cite{ute2_nat20, triplet_jpsj12, triplet1, Fay}.}
\end{abstract}

\maketitle

Symmetry plays a crucial role in determining the physical properties of solids \cite{symmetry1, symmetry2, symmetry3, Zak}. Breaking of symmetry leads to diverse physical phases and phase transitions. For example, crystalline order in solids is obtained by breaking the translational symmetry of an amorphous phase, ferromagnetism may emerge via the breaking of the rotational symmetry of a paramagnetic phase, the spontaneous breaking of gauge symmetry facilitates superfluidity and superconductivity\cite{BCS}, and breaking of the time reversal symmetry lifts Kramer's degeneracy and has close connection to a number of exotic quantum phenomena, such as quantum anomalous Hall effect\cite{QAHE, QAHE1}. 
Even more complex varieties of physical properties like ferroelectricity, spin-orbit physics, Weyl physics, unconventional superconductivity, \emph{etc.} are often obtained by breaking of the inversion symmetry\cite{broken_inversion, broken_inversion1, ncs_rev14, ncs_book12}. In a given family of compounds, keeping the chemical composition same, diverse electronic and magnetic states can be achieved simply by tuning the symmetry of the systems\cite{Graphene2}.

\begin{figure}[h!]
	\includegraphics[width=1\linewidth]{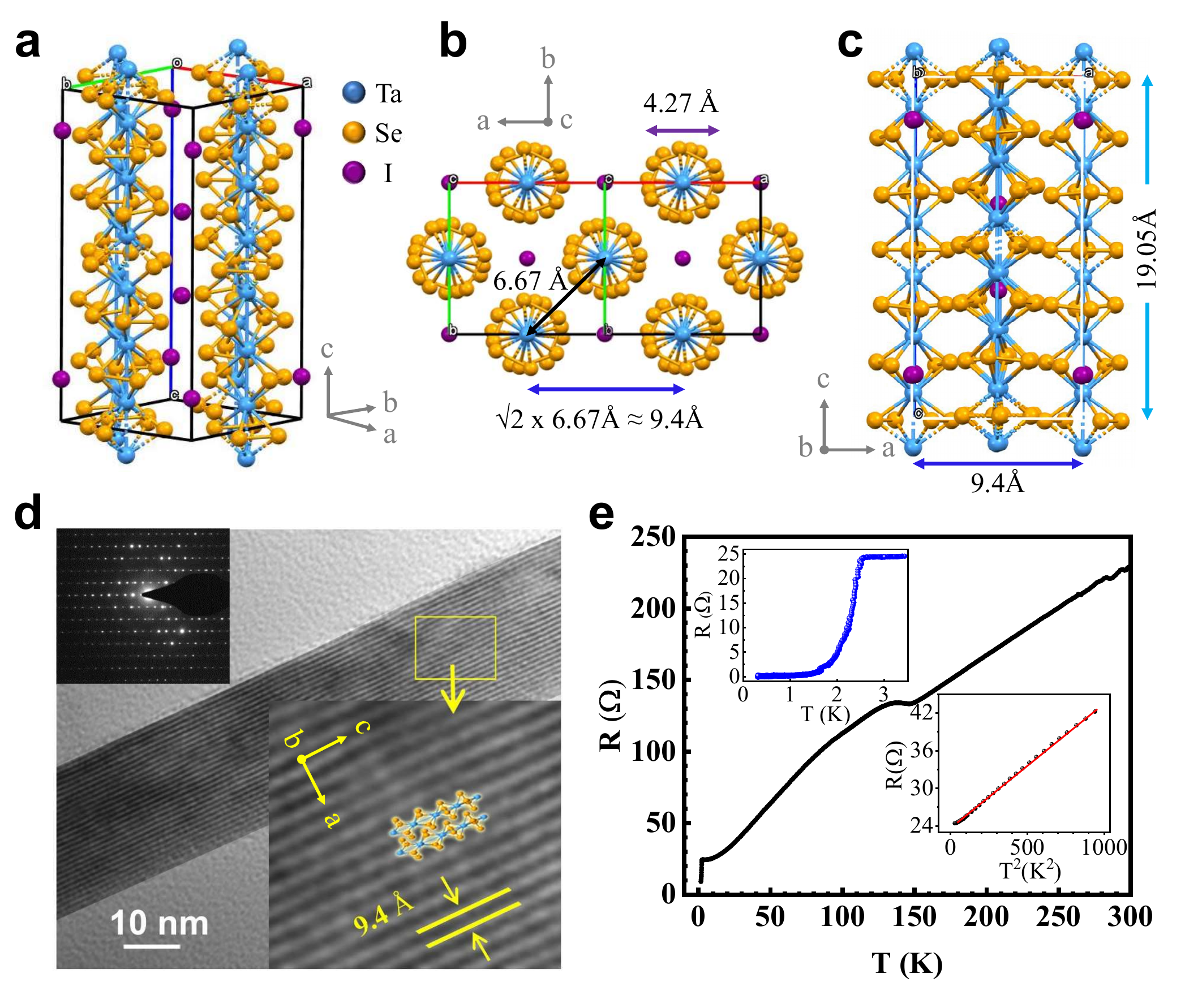}
	\caption{\textbf{Crystal structure \& superconducting transition:} Schematic of the simple tetragonal unit cell \textbf{a,} oblique view, \textbf{b,} projected view on the $ab$-plane and \textbf{c,} linear chains in lateral view. \textbf{d,} Quasi-1D structure as seen from distinct lattice chains in the high resolution transmission electron microscopy (HRTEM) image. Upper \emph{inset}: A selected area electron diffraction (SAED). Lower $inset:$ The zoomed image showing an inter-chain separation of $\approx9.4$\,\AA. \textbf{e,} Temperature dependence of resistance ($R$) measured in a conventional 4-probe geometry. The change in slope around 150\,K signifies a charge density wave (CDW) transition in $n$-TSI. Upper \emph{inset}: The superconducting transition at zero magnetic field. Lower \emph{inset}: Linear behaviour of $R$ \emph{vs.} $T^2$ in the temperature range of 5-25\,K with zero magnetic field.}
	\label{f1}
\end{figure}

The family of linear-chain compounds ($M$Se$_4$)$_n$I where $M$=Nb and Ta and $n$=2, 3, and 10/3, have attracted the attention of the community time and again because they display wide varieties of phase transitions and phase co-existences\cite{cTSI1, cTSI2, cTSI3, NbSe4_2I_prb88, TaSe4_2I_natphys21, TaSe4_2I_am20, TaSe4_2I_prb20, TaSe4_2I_nat19, TaSe4_2I_prl13}. In this family of compounds, the MSe$_4$ chains, which are separated from each other by the iodine atoms, run along the $c$ axis thereby giving rise to quasi-one-dimensional structure\cite{cTSI1, cTSI2, cTSI3, NbSe4_2I_prb88}. Tight-binding band structure calculations reveal that the electronic properties of these compounds depend on the value of $n$, which drives the electronic character by altering the degree of band filling\cite{cTSI3, NbSe4_2I_prb88}. Being quasi-one-dimensional conductors, they are prone to undergo a Peierls transition leading to a charge-density wave (CDW) phase transition at low temperature. Recently, a member of  the (MSe$_4$)$_n$I family with $n$ = 2, (TaSe$_4$)$_2$I, was shown to host an exotic axionic CDW phase\cite{TaSe4_2I_natphys21, TaSe4_2I_nat19}. However, for $n=$3, (NbSe$_4$)$_3$I and (TaSe$_4$)$_3$I are known to form a centrosymmetric structure with space group $P4/mnc$\cite{cTSI2}. Here, we have stabilized the (TaSe$_4$)$_3$I in a non-centrosymmetric phase  (space group: $P\bar{4}2_1c$) over a broad temperature range and hereafter, we refer it as $n$-TSI phase. In this phase, interestingly, the system is metallic and shows a CDW transition around 150\,K. With further lowering the temperature, it undergoes a long-range magnetic transition around 10 K and eventually settles in a superconducting ground state coexisting with ferromagnetism below 2.5\,K. This unique coexistence seemingly arises from inversion symmetry breaking as the centrosymmetric phase of the same compound (space group: $P4/mnc$) is known to be a non-magnetic insulator\cite{cTSI1, cTSI2, cTSI3, NbSe4_2I_prb88}.

 Single crystals of $n$-TSI were grown by chemical vapor transport method (refer to method section). The crystals are formed in ribbon-like fibres of length $\sim$ few mm, width $\sim$ few micron (see supplemental information for more details - Figure~S1-S4). Single crystal X-ray analysis at 100~K revealed that $n$-TSI  has a simple tetragonal unit cell (space group P$\bar{4}$2$_1$c, no. 114, CCDC entry 2055811). The lattice parameters are: $a=b=9.4358(5)\,\text{\AA}$, $c=19.0464(11)\,\text{\AA}$; $\alpha=\beta=\gamma=90^{\circ}$. The schematic representation of the structure is illustrated in Figure~\ref{f1} a-c. As shown in Figure~\ref{f1}.d, atomic resolution imaging under a transmission electron microscope revealed linear TaSe$_4$ chain-like structure along the $c$-axis. The distance between TaSe$_4$ chains is $d_{inter}=6.677\,\text{\AA}$ and the diameter of each TaSe$_4$ chain is $d_{intra}=4.271$\,\AA. 

To investigate the electronic properties of the needle-like crystals, we have employed usual four-probe resistivity measurements on a bunch of ribbons as well as on a single ribbon. As shown in Figure \ref{f1}e (multi ribbon device), Resistance (R) decreases rapidly with decreasing temperature, indicating good metallic behavior. Furthermore, the linear nature of the $R$ \emph{vs.} $T^2$ plot (in the lower inset) in the temperature range of 5-25\,K confirms that $n$-TSI is a clean Fermi liquid metal. $R$($T$)  curve shows a slope change around 150\,K which is known to arise for a charge density wave (CDW) transition due to Peierls-like instability of the linear chain structure \cite{cTSI1, cTSI2, TaSe4_2I_natphys21, TaSe4_2I_am20, TaSe4_2I_prb20, TaSe4_2I_nat19, TaSe4_2I_prl13}. Below 2.5\,K, the resistance shows a sharp fall which resembles a superconducting transition. When the samples were cooled in a dilution refrigerator with a lower base temperature, the zero-resistance state was achieved below $\sim$1K, thereby confirming superconductivity (see the upper $inset$ of Figure \ref{f1}e). In the supplementary information, Figure~S5 displays much sharper zero-resistance transition for a single ribbon device, indicating that superconductivity is indeed an intrinsic property of the quasi-one dimensional crystals.

\begin{figure}
	\includegraphics[width=0.7\linewidth]{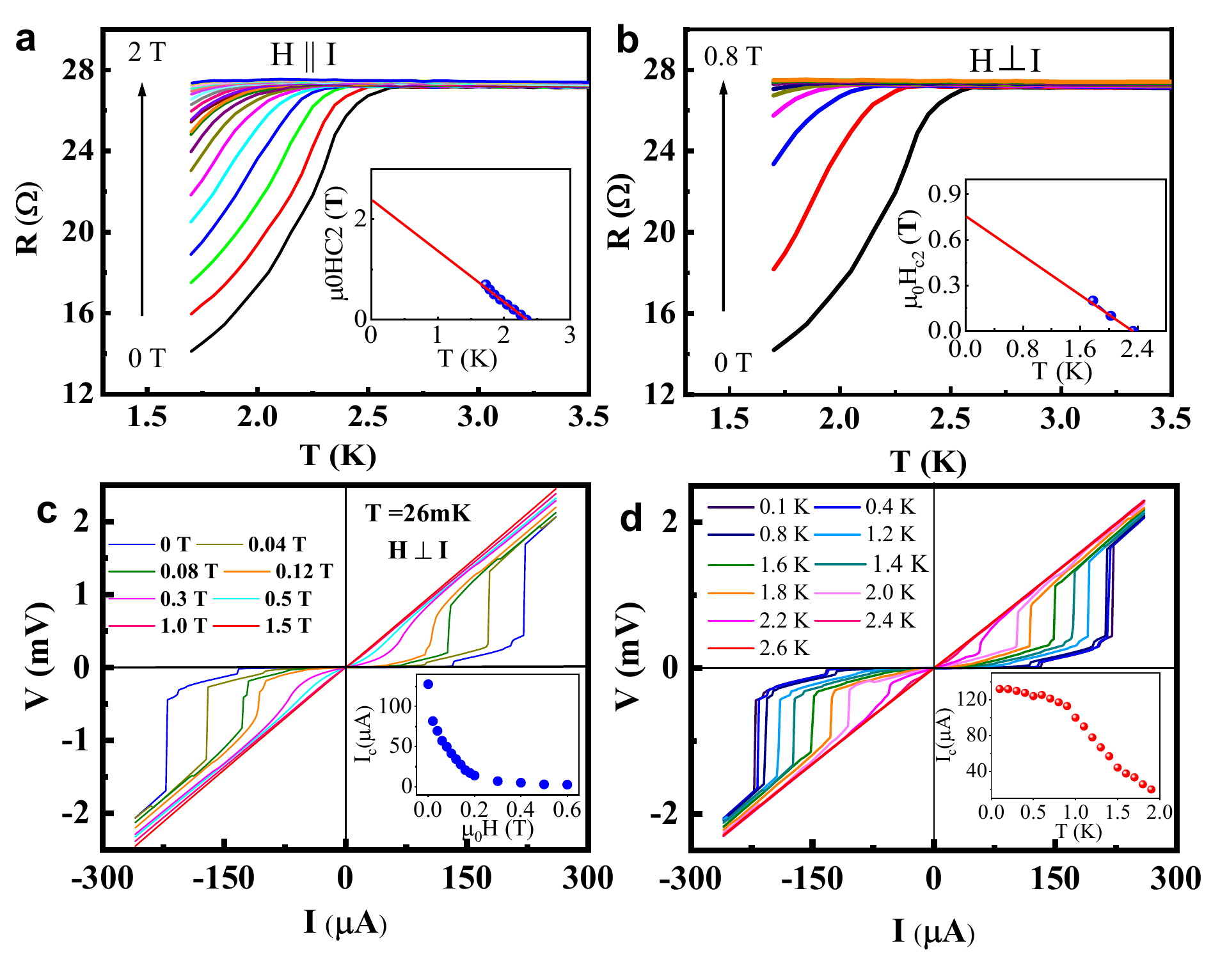}
	\caption{\textbf{Superconducting transition in magneto-transport measurements:} Systematic suppression of the superconductivity with external magnetic field applied \textbf{a,} along the direction of current which is parallel to the $c$-axis ($H||$c) and \textbf{b,} perpendicular to the direction current \emph{i.e.} $H\perp$c. H$_{c2}$ vs. T plots along both directions are shown in the respective \emph{inset}. \textbf{c,} Critical-current dominated $I-V$ characteristic at 26\,mK with different magnetic fields (above and below $H_{c2}\perp c$). The magnetic field variation of critical current ($I_c$) is shown in the \emph{inset}. \textbf{d,} Zero-field $I-V$ characteristic at different temperatures and temperature dependence of critical current ($I_c$) in the \emph{inset}.}
	\label{f2}
\end{figure}

With the application of a magnetic field parallel to the length of the needle-like crystals ($H||I||c$), the transition temperature ($T_c$) gradually decreases and goes beyond our measurement limit above a magnetic field of $H^{||c}_{c2}\simeq2$\,Tesla, as shown in Figure \ref{f2}a. This is expected for a superconducting phase transition. The zero-temperature critical field, $H^{||c}_{c2}(0)\simeq2.35$\,Tesla is estimated from an approximated linear extrapolation (this is approximate, as there is a weak non-linearity in $H^{||c}_{c2}$ \emph{vs.} $T_c$ data ($inset$ of Figure \ref{f2}a). Here, $T_c$ has been taken as the temperature where the resistance drops to 90\% of the normal state resistance. Possibly due to an anisotropy in the superconducting state, originating either from the geometry of the crystals or from an intrinsic anisotropy, the critical field is observed to be smaller when the magnetic field is applied perpendicular to the $c$-axis. The estimated zero-temperature critical field in this case is $H^{\perp c}_{c2}(0)\simeq0.75$\,Tesla (Figure \ref{f2}b). 


To further investigate the transport properties, we have recorded the $I-V$ characteristics in the standard four-probe geometry down to 26\,mK. The $I-V$ curves are found to be non-linear with a typical shape dominated by critical current. A sharp jump in voltage is seen at a critical current ($I_c$). $I_c$ goes down with increasing magnetic field perpendicular to current (see the $inset$ of Figure \ref{f2}c), as expected for critical-current dominated $I-V$ characteristics. We also found that the critical current dominated feature disappears at  temperature above 2.5\,K (see Figure \ref{f2}d). All the above observations collectively confirm beyond any ambiguity that the resistive transition discussed above is due to superconductivity.

\begin{figure}
  \centering
  	\includegraphics[width=1\linewidth]{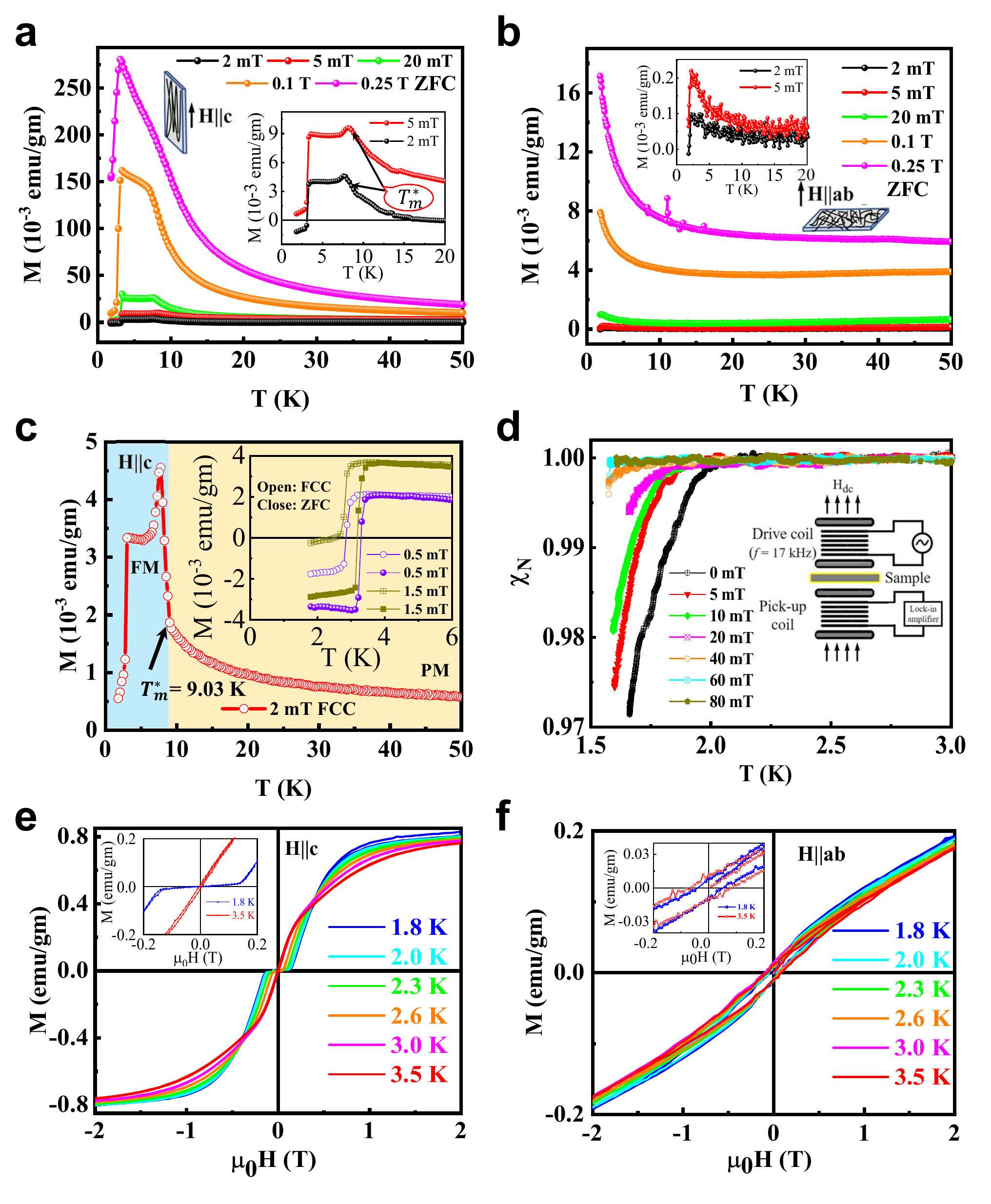}
\end{figure}
\clearpage
\begin{figure}
  \caption{\textbf{Ferromagnetism and Meissner effect:} Zero-field-cooled (ZFC) magnetization ($M-T$) curves with magnetic field applied \textbf{a,} parallel ($H||c$)and \textbf{b,} perpendicular ($H||ab$) to $c$-axis with low-field behavior in respective \emph{inset}. \textbf{c,} Field-cooled (FC) magnetization curve with $H||c$ shows the paramagnetic to ferromagnetic phase transition of n-TSI. \emph{Inset}: Temperature-dependent  magnetization data at very low applied magnetic field at 0.5 mT and 1.5 mT showing the Meissner state below 3.0 K. \textbf{d,} Two-coil mutual inductance measurements show systematic change in pick-up signal of the secondary coil as a function of dc magnetic field due to screening in the superconducting phase. \emph{Inset}: Schematic of the two-coil mutual inductance measurement set-up. \textbf{e,} Hysteresis loops with $H||c$ at different temperatures above and below $T_c$. Well above $T_c$, 
  \emph{Inset}: M-H curves obtained above and below $T_c$ (1.8\,K and 3.5\,K, respectively) show strong diamagnetic contribution below $T_c$. \textbf{f,} Hysteresis loops with $H||ab$ at different temperatures above and below $T_c$, confirming ferromagnetic ordering.}
  \label{f3}
\end{figure}
\clearpage

As per the general understanding of the microscopic theory of superconductivity, due to the breaking of $U(1)$ symmetry in the superconducting phase, superconductors are also expected to be strongly diamagnetic. In order to probe the magnetic state of $n$-TSI, detailed $dc$ magnetization measurements were performed. As shown in Figure \ref{f3}a, when the magnetic field is applied along the length of the needles ($c$-axis), magnetization ($M$) in zero-field-cooled (ZFC) cycle monotonically increases with decreasing temperature and displays a peak around 9\,K. Such a peak is often seen in systems with a helimagnetic order~\cite{heli, mnsi_uspekhi11}, where a ferromagnetic component of the order is also expected. To gain further insight on the nature of magnetic ground state, the paramagnetic susceptibility has been fitted to the Curie-Weiss (CW) expression as shown in supplementary information (Figure S12). The fit yields  positive CW temperature ($T_{CW}\sim$ 5.5 K) which suggests that magnetism in $n$-TSI is dominated by ferromagnetic interactions. 

To note, when the magnetic field is applied perpendicular to the $c$-axis (Figure \ref{f3}b), $M$ is found to be significantly smaller and the anomaly due to magnetic transition is not clearly visible. This indicates that the hard magnetic axis is perpendicular to the $c$-axis (see Section~S3, supplementary information for additional data regarding anisotropic behavior). The $inset$ of Figure \ref{f3}a shows that $M$ drops sharply below 3 K and becomes negative (diamagnetic) for low excitation field (~2\,mT). Sharp drop in $M$ has also been observed at higher fields. The diamagnetic signal is also observed when the field is applied perpendicular to the $c$-axis ($inset$ of Figure \ref{f3}b). Motivated by this, we attempted to detect the Meissner signal which should appear in the field-cooled (FC) mode. As shown in Figure \ref{f3}c, at 2\,mT, magnetization decreases sharply below 3 K, confirming strong diamagnetic contribution.  However, a clear diamagnetic signal in FC mode is observed at low temperature for lower fields (inset of Figure~\ref{f3}c). This is remarkable because in ferromagnetic superconductors, the diamagnetic signal was never directly observed before~\cite{triplet_sc19,Saxena,Aoki,Huy,zrzn2_nat01,U_SC19}. Further, to detect the shielding effect, we have measured dc-field-dependent high-frequency susceptibility of another small piece of $n$-TSI using a two-coil method. As shown in Figure 3d, the superconducting transition along with its systematic field dependence was clearly observed that confirms the superconducting ground state of $n$-TSI. It may also be mentioned that the peak at 9 K becomes much sharper in FC mode (Figure~\ref{f3}c).

\begin{figure}[h!]
	\includegraphics[width=1\textwidth]{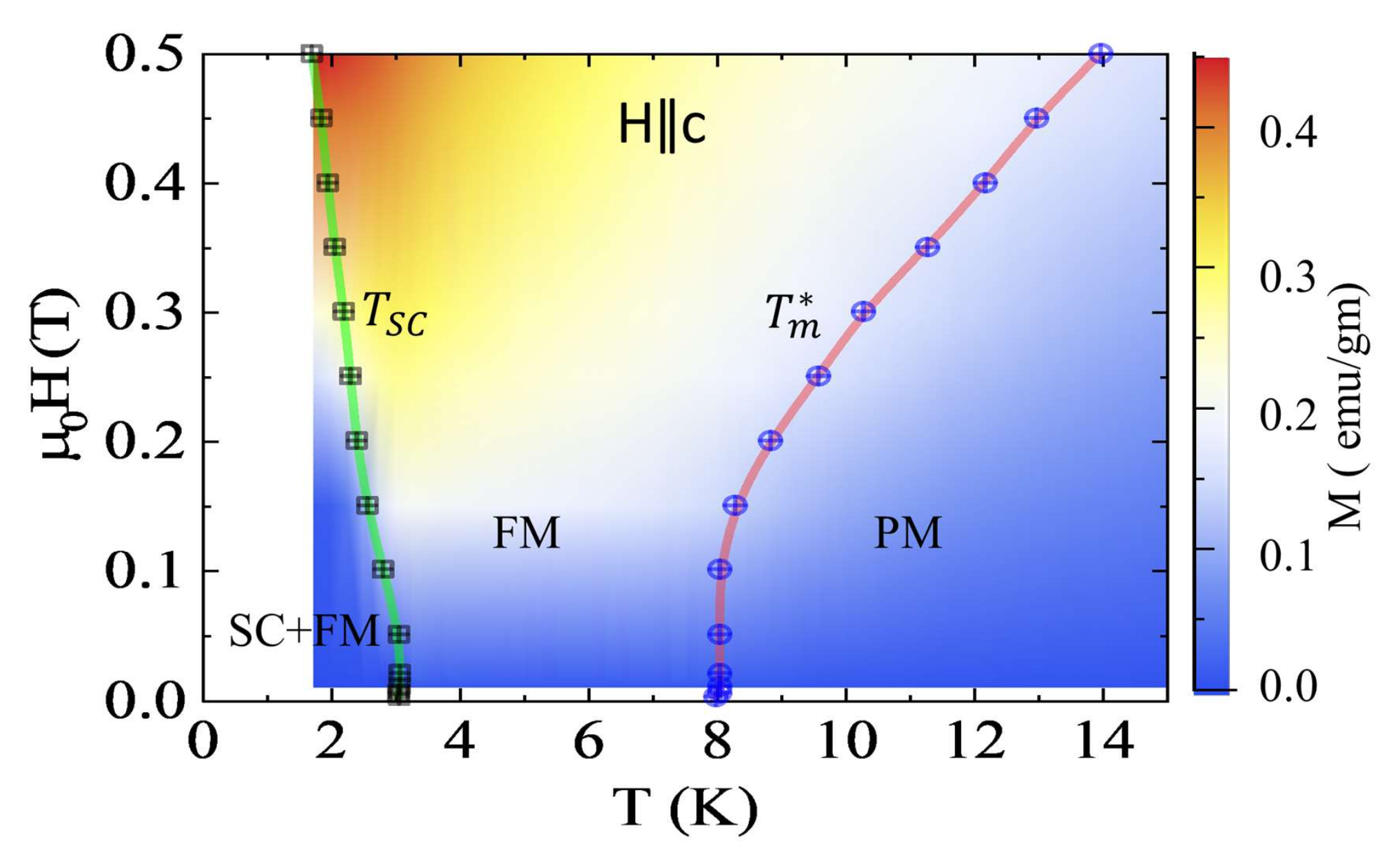}
	\caption{\textbf{Phase diagram - Superconductivity and Ferromagnetism:}  Phase diagram of ferromagnetic superconductor, nTSI based on ZFC magnetization vs. temperature (M-T) data with the applied magnetic field parallel to $c$-axis. The superconducting (SC) transition temperature ($T_{SC}$) is estimated when M shows sudden drop due to Meissner effect (see supplementary). The paramagnetic (PM) to ferromagnetic (FM) phase transition temperature ($T^*_{m}$) is defined where M shows the signature of magnetic ordering i.e. peak in M-T data (or dip in the dM/dT vs T plot as shown in supplementary).}
	\label{f4}
\end{figure}

Now we focus on the temperature dependence of $M-H$ hysteresis loops across the superconducting transition. Figure \ref{f3}e shows a set of four-quadrant $M-H$ loops below and above  $T_c$, for the field $H$ applied along the $c$-axis of the needle. A clear $S$-shaped loop is observed albeit with an extremely small coercive field at temperatures 3.0\,K and above, that is the signature of a soft ferromagnetic phase. Below $T_c$, the loops deviate from an ideal $S$ shape and show a flatter region at low fields. This change in nature can be understood for a superconducting phase coexisting with ferromagnetism. At low fields, the diamagnetic signal due to superconductivity competes with the ferromagnetic background and thereby makes the total moment of the system extremely small. At higher fields, when superconductivity is  suppressed, the ferromagnetic signal dominates and $M$ shows an upward jump. The length of the flat region in the $M-H$ loop increases with decreasing temperature, reflecting monotonic increase in $H_{c2}$ with decreasing temperature. Therefore, the $M-H$ data confirm that the superconducting phase coexists with ferromagnetism. While applying the magnetic field along the direction perpendicular to the $c$-axis (Figure 3f, raw magnetization data before diamagnetic correction is presented in Section~S4, Supplementary Information), a prominent hysteresis loop is seen across the superconducting $T_c$ for all temperatures between 1.8\,K and 3.5\,K. However, in this case, no flattening of the $M-H$ curve is observed below the superconducting $T_c$, depicting that both magnetic and superconducting properties of $n$-TSI are strongly anisotropic in nature.

Therefore, we have shown that $n$-TSI is metallic at high temperatures and undergoes a ferromagnetic transition below 10\,K and superconducting transition below 2.5\,K. All these observations can be summarized in a phase diagram, deduced directly from the experimental data, as shown in Figure~\ref{f4}. The presence of a ferromagnetically ordered state in $n$-TSI is surprising mainly because the material does not host any magnetic element. 
The observed weak ferromagnetic state~\cite{fm_rmp16, spin fluctuation_book_moriya} could be a consequence of spin-orbit coupling which appears ubiquitously in inversion-symmetry broken materials. Furthermore, the fact that the system undergoes multiple phase transitions just above the superconducting critical temperature hints to the possibility of a quantum critical state\cite{Grigera, Mackenzie, Kirkpatrick, Ovchinnikov, hf_rmp09}. Finally, coexistence of ferromagnetism and superconductivity with the indication of anisotropic nature of superconductivity that we found here, warrants further research for probing the symmetry of the order parameter in $n$-TSI. Such a system is ideal for the existence of the hitherto elusive spin-triplet pairing symmetry\cite{ute2_nat20, triplet_jpsj12, triplet1, Fay}.

\section{Methods:}

\subsection{Sample Growth:} n-TSI single crystals were grown by chemical vapor transport (CVT) method. Ta (Alfa Aeser, purity 99.97\%) and Se (Alfa Aeser, purity 99.999\%)  in powder form were mixed with iodine pieces (Alfa Aeser, 99.0\%) in 2.85:4.977:1 (Ta:Se:I) weight ratio and were ground together into fine powder. The powder was loaded into a quartz tube and vacuum sealed at 10$^{-5}$\,mbar. The tube was placed in a two-zone furnace for one week maintained at 500$^{\circ}$\,C and 400$^{\circ}$\,C. Hair/fibre like single crystals were grown along random directions. After cooling down to room temperature, the crystals (silver/grey in color) were collected as lump of fibre wool by breaking the quartz tube in air at room temperature ($\sim$300\,K). The material is stable under ambient condition.
\subsection{Characterization \& Measurements:} Single crystal XRD characterization were done on a Bruker D8 VENTURE Microfocus diffractometer equipped with PHOTON II Detector, with Mo K$_{\alpha}$ radiation ($\lambda=0.71073$\,\AA) at 100 K. The transmission electron microscopy (TEM) studies were performed using a FEI, TECNAI G2 F30, S-Twin microscope. Electrical transport and magneto-transport measurements were performed in a commercial physical property measurement system (PPMS, Quantum Design) and sub-Kelvin measurements were carried out in a dilution refrigerator (Oxford Instrument). Electrical connections were made in 4-probe geometry using silver-paint (RS components). DC magnetization measurements were performed in a squid magnetometer (SQUID-VSM Evercool, MPMS 3 from Quantum Design). Two coil mutual inductance measurements were performed by a home-made two coil set-up in a He-4 wet cryostat equipped with a 3-axis vector magnet (6T-1T-1T).

\section{Acknowledgements:} The work at IACS and SINP was supported by the Department of Science and Technology under SERB research grant (Grant No. SRG/2019/000674, ECR/2017/002037 and EMR/2016/005437). MM and AB acknowledges financial support from CSIR (CSIR Grant No. 09/080(1109)/2019-EMR-I respectively). Authors also would like to acknowledge Partha Rana and Tanmoy Das for fruitful discussions. GS acknowledges financial support from the Department of Science and Technology (DST), Govt. of India through Swarnajayanti fellowship (grant number: DST/SJF/PSA-01/2015-16).

\section{Author contributions:} MM and AB conceived the project. AB, SG, and MM planned and coordinated experiments. AB and SG have synthesized the single crystal and did the initial characterization and measurements under the supervision of MM. AB, BS, BD, MJ and RAS performed single crystal XRD, XPS, TEM and SEM. SM did the magnetization measurements and analyzed the data under the guidance of PM. BP and RP fabricated devices and performed detailed transport measurements down to 1.7 K and TM did the ultra-low temperature transport measurements in dilution refrigerator. Whole transport measurements and analysis were carried out under the guidance of ANP. AV, SH and GS performed the two-coil measurements. GS wrote the manuscript with the help of other co-authors.


\begin{thebibliography}{100}


\bibitem{cTSI1} Gressier, P. \emph{et al.} Characterization of the new series of quasi one-dimensional compounds (MX$_4$)$_n$Y (M = Nb, Ta; X = S, Se; Y = Br, I). \emph{J. of Solid State Chem.} \textbf{51}, 141-151 (1984).

\bibitem{cTSI2} Roucau, C. \emph{et al.} Electron microscopy study of transition-metal tetrachalcogenide (MSe$_4$)$_n$I (M = Nb, Ta). \emph{J. Phys. C: Solid State Phys.} \textbf{17}, 2993 (1984).

\bibitem{cTSI3} Gressier, P. \emph{et al.} Electronic structures of transition-metal tetrachalcogenides (MSe$_4$)$_n$I (M = Nb, Ta). \emph{Inorg. Chem.} \textbf{23}, 1221 (1984).

\bibitem{NbSe4_2I_prb88} Sekine, T. \& Izumi, M. Successive phase transitions in the linear-chain semiconductor (NbSe$_4$)$_3$I studied
by Raman scattering and electrical resistivity. \emph{Phys. Rev. B} \textbf{38}, 2012 (1988).


\bibitem{ute2_nat20} Jiao, L. \emph{et al.} Chiral superconductivity in heavy-fermion metal UTe$_2$. \emph{Nature} \textbf{579}, 523527 (2020).

\bibitem{triplet_jpsj12} Maeno, Y. \emph{et al.} Evaluation of Spin-Triplet Superconductivity in Sr$_2$RuO$_4$. \emph{J. Phys. Soc. Jpn.} \textbf{81}, 011009 (2012).

\bibitem{triplet1} Mackenzie, A. P. \& Maeno, Y. The superconductivity of Sr$_2$RuO$_4$ and the physics of spin-triplet pairing. \emph{Rev. Mod. Phys.} \textbf{75}, 657(2003).

\bibitem{Fay}Fay, D. \& Appel, J. Coexistence of p-state superconductivity and itinerant ferromagnetism. \emph{Phys. Rev. B} \textbf{22}, 3173 (1980).



\bibitem{symmetry1} Lederman L. M. \& Hill, T. H. Symmetry and the Beautiful Universe. Prometheus Books (2007).

\bibitem{symmetry2} Brading, K. \& Castellani, E. (\emph{eds.}) Symmetries in Physics: Philosophical Reflections. Cambridge University Press (2003).

\bibitem{symmetry3} Gross, D. J. The role of symmetry in fundamental physics. \emph{Proc. Natl. Acad. Sci. USA} \textbf{93}, pp. 14256 (1996).

\bibitem{Zak} Zak, J. Symmetry criterion for surface states in solids. \emph{Phys. Rev. B} \textbf{32}, 2218 (1985).

\bibitem{BCS} Bardeen, J., Cooper, L. N. \& Schrieffer, J. R., Theory of Superconductivity. \emph{Phys. Rev.} \textbf{108}, 1175 (1957).

\bibitem{QAHE} Chang, C-Z. \emph{et al.} Experimental Observation of the quantum Anomalous Hall Effect in a Magnetic Topological Insulator. \emph{Science} \textbf{340} 167 (2013).

\bibitem{QAHE1} Machida, Y. \emph{et al.} Time-reversal symmetry breaking and spontaneous Hall effect without magnetic dipole order. \emph{Nature} \textbf{463}, 210, (2010).




\bibitem{broken_inversion} Torre, A. \emph{et al.} Mirror symmetry breaking in a model insulating cuprate. \emph{Nature Physics} \textbf{17}, 777 (2021).
\bibitem{broken_inversion1} Shi, Y. \emph{et al.}. A ferroelectric-like structural transition in a metal. \emph{Nature Mater.} \textbf{12}, 1024 (2013).

\bibitem{ncs_rev14} Yip, S. Noncentrosymmetric superconductors. \emph{Annu. Rev. Condens. Matter Phys.} \textbf{5}, 1517 (2014).

\bibitem{ncs_book12} Bauer, E. \& Sigrist, M. (\emph{eds.}) Non-Centrosymmetric Superconductors: Introduction and Overview.
Springer Heidelberg (2012).

\bibitem{Graphene2} Wang, E. \emph{et al.} Gaps induced by inversion symmetry breaking and second-generation Dirac cones in graphene/hexagonal boron nitride. \emph{Nature Physics} \textbf{12}, 1111, (2016).


\bibitem{TaSe4_2I_natphys21} Shi, W. \emph{et al.} A charge-density-wave Weyl semimetal. \emph{Nature Physics} \textbf{17}, 381177 (2021).

\bibitem{TaSe4_2I_nat19} Gooth, J. \emph{et. al.} Axionic charge-density wave in the Weyl semimetal (TaSe$_4$)$_2$I. \emph{Nature} \textbf{575},  315-19, (2019).

\bibitem{TaSe4_2I_am20} An, C. \emph{et. al.} Long-range ordered amorphous atomic chains as building blocks of a superconducting quasi-one-dimensional crystal. \emph{Adv. Mater.} \textbf{32}, 2002352 (2020).

\bibitem{TaSe4_2I_prb20} Zhang, Y. \emph{et al.} First-principles study of the low-temperature charge density wave phase in the quasi-one-dimensional Weyl chiral compound (TaSe$_4$)$_2$I. \emph{Phys. Rev. B} \textbf{101}, 174106 (2020).

\bibitem{TaSe4_2I_prl13} Tournier-Colletta, C. \emph{et al.} Electronic instability in a zero-gap semiconductor: the charge-density wave in (TaSe$_4$)$_2$I. \emph{Phys. Rev. Lett.} \textbf{110}, 236401 (2013).


\bibitem{pauli_para_prl62} Clogston, A. M. Upper limit for the critical field in hard superconductors. \emph{Phys. Rev. Lett.} \textbf{9}, 266 (1962).

\bibitem{pauli_para_apl62} Chandrasekhar, B. S. \emph{Appl. Phys. Lett.} \textbf{1}, 7 (1962).

\bibitem{paulipara_prl04} Frigeri, P. A. \emph{et al.} Superconductivity without Inversion Symmetry: MnSi versus CePt$_3$Si. \emph{Phys. Rev. Lett.} \textbf{92}, 097001 (2004).


\bibitem{heli} Sirica, N. \emph{et al.} The nature of ferromagnetism in the chiral helimagnet Cr$_{1/3}$NbS$_2$. \emph{Commun. Phys.} \textbf{3}, 65 (2020).

\bibitem{mnsi_uspekhi11} Stishov, S. M. \& Petrova, A. E. Itinerant helimagnet MnSi. \emph{Physics-Uspekhi} \textbf{54}, 1117 (2011).


\bibitem{triplet_sc19} Ran, S. \emph{et al.} Nearly ferromagnetic spin-triplet superconductivity. \emph{Science} \textbf{365}, 684-687 (2019).

\bibitem{Saxena} Saxena, S. S. \emph{et. al.} Superconductivity on the border of itinerant-electron ferromagnetism in
UGe$_2$. \emph{Nature} \textbf{406}, 587-92, (2000).

\bibitem{Aoki} Aoki, D. \emph{et al.} Coexistence of superconductivity and ferromagnetism in URhGe. \emph{Nature} \textbf{413}, 613176 (2001).

\bibitem{Huy} Huy, N. T. \emph{et al.} Superconductivity on the border of weak itinerant ferromagnetism in UCoGe. \emph{Phys. Rev. Lett.} \textbf{99}, 067006 (2007).

\bibitem{zrzn2_nat01} Pfleiderer, C. \emph{et al.} Coexistence of superconductivity and ferromagnetism in the d-bandmetal ZrZn$_2$. \emph{Nature} \textbf{412}, 58 (2001).

\bibitem{U_SC19} Aoki, D., Ishida, K. \& Flouquet, J. Review of U-based ferromagnetic superconductors: comparison between UGe$_2$, URhGe, and UCoGe. \emph{J. Phys. Soc. Jpn.} \textbf{88}, 022001 (2019).


\bibitem{mnsi_nw_acsnano10} Seo, K. \emph{et al.} Itinerant Helimagnetic Single-Crystalline MnSi Nanowires. \emph{ACSNano} \textbf{4}, 2569 (2010).

\bibitem{heli1}Cao, Y. \emph{et al.} Overview and advances in a layered chiral helimagnet Cr$_1/3$NbS$_2$. \emph{Mater. Today Adv.} \textbf{7}, 100080 (2020).

\bibitem{heli2}Kishine, J. \& Ovchinnikov, A. S. Theory of monoaxial chiral helimagnet in: \emph{Solid State Physics} (R. E. Camley and R. L. Stamps, eds.), Acad. Press, New York \textbf{66}, 1170 (2015).


\bibitem{fm_rmp16} Brando, M. \emph{et al.} Metallic quantum ferromagnets. \emph{Rev. Mod. Phys.} \textbf{88}, 025006 (2016).

\bibitem{spin fluctuation_book_moriya} T. Moriya, Spin Fluctuations in Itinerant Electron Magnetism (Springer, Berlin, 1985).



\bibitem{Grigera} Grigera, S. A. \emph{et al.} Magnetic field-tuned quantum criticality in the metallic ruthenate Sr$_3$Ru$_2$O$_7$. \emph{Science} \textbf{294}, 329 (2001).

\bibitem{Mackenzie} Mackenzie, A. P. \& Grigera, S. A. A Quantum Critical Route to Field-Induced Superconductivity. \emph{Science} \textbf{309}, 1330 (2005).

\bibitem{Kirkpatrick} Kirkpatrick, T. R. \& Belitz, D. Ferromagnetic Quantum Critical Point in Noncentrosymmetric Systems. \emph{Phys. Rev. Lett.} \textbf{124}, 147201 (2020).

\bibitem{Ovchinnikov} Ovchinnikov, A. S. \emph{et al.} Critical behavior of a monoaxial chiral helimagnet. \emph{Theo. Math. Phys.} \textbf{191}, 924 (2017).

\bibitem{hf_rmp09} Pfleiderer, C. Superconducting phases of $f$-electron compounds. Reviews of Modern Physics. \emph{Rev. Mod. Phys.} \textbf{81}, 1551 (2009).




%
















\end{thebibliography}
\end{document}